\documentclass{article}

\usepackage{arxiv}

\usepackage[utf8]{inputenc} 
\usepackage[T1]{fontenc}    
\usepackage{hyperref}       
\usepackage{url}            
\usepackage{booktabs}       
\usepackage{amsfonts}       
\usepackage{nicefrac}       
\usepackage{microtype}      
\usepackage{lipsum}		
\usepackage{graphicx}
\usepackage[numbers]{natbib}
\usepackage{doi}

\usepackage{cite}
\usepackage{amsmath,amssymb,amsfonts}
\usepackage{graphicx}
\usepackage{textcomp}
\usepackage{multirow}
\usepackage{times}
\usepackage{amsmath,nccmath}
\usepackage{cuted}
\usepackage{algpseudocode}
\usepackage{algorithm}
\usepackage{caption}
\usepackage{subcaption}
\usepackage{array}
\title{Mental Stress Detection and Performance Enhancement Using FNIRS and Wrist Vibrator Biofeedback}

\date{} 					

\author{ \href{https://orcid.org/0009-0002-7450-3113}{\hspace{1mm}Anita Beigzadeh} \\
	Department of Electrical Engineering\\
	Tehran University\\
	\texttt{anitabeigzadeh@ut.ac.ir} \\
	\And
	\href{https://orcid.org/0009-0004-4911-9624}{\hspace{1mm}Vahid Yazdnian} \\
	Department of Electrical Engineering\\
	Princeton University\\
	\texttt{vahidyazdnian@princeton.edu} \\
        \And
	\href{https://orcid.org/0000-0001-5928-8668}{\hspace{1mm}Seyed Kamaledin Setarehdan} \\
	Department of Electrical Engineering\\
	Tehran University\\
	\texttt{ksetareh@ut.ac.ir} \\
}

\renewcommand{\shorttitle}{Mental Stress Detection and Performance Enhancement}

\hypersetup{
pdftitle={A template for the arxiv style},
pdfsubject={q-bio.NC, q-bio.QM},
pdfauthor={David S.~Hippocampus, Elias D.~Striatum},
pdfkeywords={First keyword, Second keyword, More},
}

\begin{document}
\maketitle

\begin{abstract}
	Any person in his/her daily life activities experiences different kinds and various amounts of mental stress which has a destructive effect on their performance. Therefore, it is crucial to come up with a systematic way of stress management and performance enhancement. This paper presents a comprehensive portable and real-time biofeedback system that aims at boosting stress management and consequently performance enhancement. For this purpose, a real-time brain signal acquisition device, a wireless vibration biofeedback device, and a software-defined program for stress level classification have been developed. More importantly, the entire system has been designed to present minimum time delay by propitiously bridging all the essential parts of the system together. We have presented different signal processing and feature extraction techniques for an online stress detection application. Accordingly, by testing the stress classification section of the system, an accuracy of 83\% and a recall detecting the true mental stress level of 92\% was achieved. Moreover, the biofeedback system as integrity has been tested on 20 participants in the controlled experimental setup. Experiment evaluations show promising results of system performances, and the findings reveal that our system is able to help the participants reduce their stress level by 55\% and increase their accuracy by 24.5\%. It can be concluded from the observations that all primary premises on stress management and performance enhancement through reward learning are valid as well.
\end{abstract}


\section{Introduction}
\label{sec:introduction}
Stressful activities in daily life have always been a serious health problem for people working under high amounts of mental pressure. Examples are students dealing with their exams, professionals with a large number of work responsibilities, employees who work in a prestigious company and are under a great workload, and/or a businessman who risks all their capital. While dealing with these activities is inevitable, many serious side effects can be caused due to high mental pressure on these people in a life period. Stress can be identified as a state of worry or mental tension caused by a difficult situation. Several studies have been carried out to show the devastating health problems of stress. It has been shown that doing daily stressful activities can severely influence the well-being and state of health of any person \citep{delongis1988impact,mcewen2008central,anderson1998levels}. Through examinations, it has been proved that there is a strong relationship between high daily stress levels and the occurrence of concurrent and subsequent health problems such as flu, sore throat, headache, and backaches or even much more serious problems like lack of the immune system, cardiovascular system, neuroendocrine system, and central nervous systems \citep{delongis1988impact,anderson1998levels}. Furthermore, although the effects of stress on mood changes are complex, it has shown to be very influential \citep{delongis1988impact}. Therefore, participating in daily stressful activities can cause serious health problems and diseases along with a great impact on performance and procrastination. In \citep{shankar2016effects}, it was shown that students’ physical and mental health are affected by stress. Statistical analysis shows that nearly 50\% of the students blame schools for being a substantial source of stress causing performance reduction and suffering from the side effects of stress (World Health Organization) \citep{shankar2016effects,coffman2002social}.\newline
As previously mentioned people from different backgrounds are struggling with the catastrophic effects of stress in the long term in various ways caused by daily stressful activities. To counteract the negative effects of stress on mental and physical health along with performance diminution of individuals one should measure the stress level at the very first step so as to take that information into account and act properly. In general, there are various invasive and non-invasive techniques for stress level measurement. \citep{marques2010evaluation,makra2021psychological}. The very basic and classical method for stress level assessment is via blood cortisol level measurement which is an invasive method and it has been shown that there is a strong correlation between the stress level and the blood cortisol level \citep{lee2015technical}. Although, there have been other studies considering urinary, salivary, sweat, and tears cortisol level assessment methods for stress level measurement \citep{lee2015technical,hellhammer2009salivary,elias2014late}, they are all considered as invasive techniques in our context for using stress level measurement to deal with daily stressful activities. In contrast, non-invasive methods use biometric signals and features such as brain signals, heart signals, motion signals, and thermographic signals to assess the stress level which is very suitable for real-time stress measurement applications since as opposed to invasive methods they can assess the stress level fast so they are so suitable for biofeedback applications \citep{marques2010evaluation,al2016mental,gunawardhane2013non}.\newline
 Brain signals have the most significant correlation with mental stress level and the majority of research studies in stress level assessment have utilized brain signals. Therefore, it is reasonable to conclude that brain signal features are of much higher importance for measuring the stress level in people. Generally, there are two popular methods to record the brain signals namely Functional Near Infrared Spectroscopy and Electroencephalogram which both can be used with different channel configurations on the scalp. Specifically, research studies incorporate EEG and fNIRS brain signal recording for stress assessment \citep{al2017stress,al2015simultaneous}. Fundamentally, the recorded signals in previous studies would be processed and evaluated using various techniques with different objectives and the post-processed signal then would be utilized in different classification models to evaluate the stress level. There are typically two types of classification that were previously used for this purpose which include Linear and non-linear feature extraction techniques along with machine learning classification, and Deep Convolutional Neural Networks.
Regarding the first method, the principle is using novel feature extraction techniques on recorded brain signals and incorporating the gathered features to firstly train the classification model that can be KNN, SVM, MLP, etc., and secondly using them in the test phase to evaluate the stress level \citep{arefi2018classification,hakimi2018stress}. This method is fast and simple yet efficient to classify the stress level, but it requires proper and sufficient levels of signal processing and feature extraction. The other methodology uses deep neural networks; while they can be deployed easily without prior processing, they need a large amount of data for training as the model grows and gets complicated. Another disadvantage of this method is that stress level classification is time-consuming because of the intrinsic complexity of deep neural networks which makes them inappropriate for real-time applications. There are research studies incorporating CNN, RNN, and LSTM for stress assessment using brain signals \citep{sundaresan2021evaluating,hakimi2020proposing}.\newline
A comprehensive evaluation of stress levels serves as a pivotal indicator for assessing one's mental health. It is not merely sufficient to recognize the presence of stress; it is equally imperative to implement tailored interventions that align with the identified stress levels, effectively mitigating the adverse impacts of mental anxiety. This paper is dedicated to the main objective of counteracting elevated mental stress and subsequently diminishing its influence to bolster overall performance. While various techniques have been explored in the existing literature for stress management, our focus in this article gravitates toward the innovative Biofeedback methodology, as highlighted in previous studies \citep{chiesa2009mindfulness,purwandini2012resonant,de1980cognitive}. Biofeedback has been previously used in auditory and visualized forms for various conceptual purposes \citep{spada2022heart,hunter2019effect,brinkmann2020comparing}, but there is no limitation for biofeedback forms as it can be used also in vibration and commendatory forms. In addition, biofeedback, characterized as a feedback loop that heightens individuals' awareness of crucial bio-indicators, transcends its traditional applications and finds its place in human-computer interaction systems as well. In a study, the incorporation of motion signals in conceptual and gamified feedback forms proved biofeedback to be beneficial for rehabilitation purposes \citep{yazdnian2022comprehensive}.

\begin{figure}[t]
    \centering
    \begin{minipage}{0.45\linewidth}
        \centering
        \includegraphics[width=\linewidth, height=6cm]{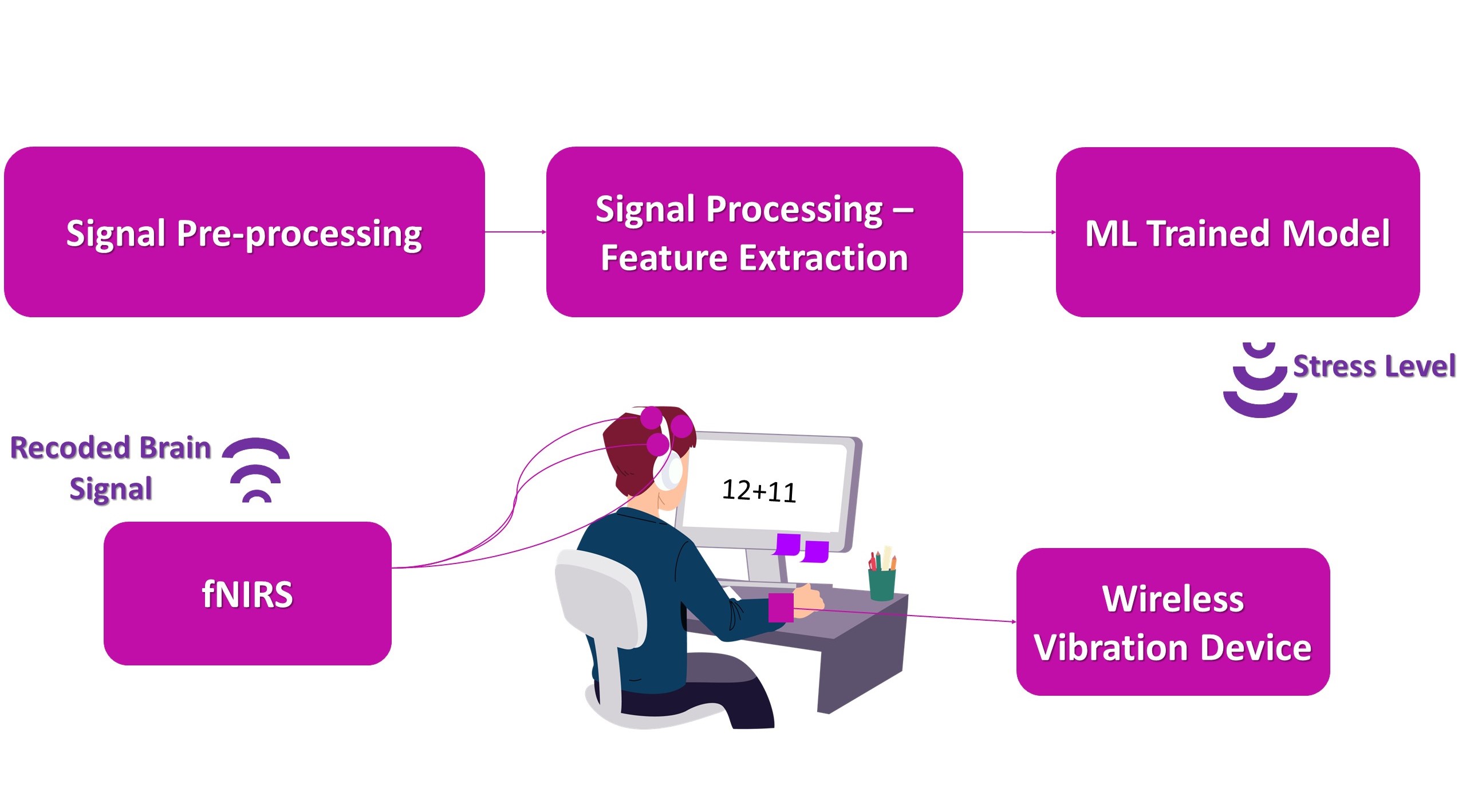}
        \caption{Comprehensive Biofeedback System}
        \label{fig:0}
    \end{minipage}%
    \hfill
    \begin{minipage}{0.45\linewidth}
        \centering
        \includegraphics[width=\linewidth, height=6cm]{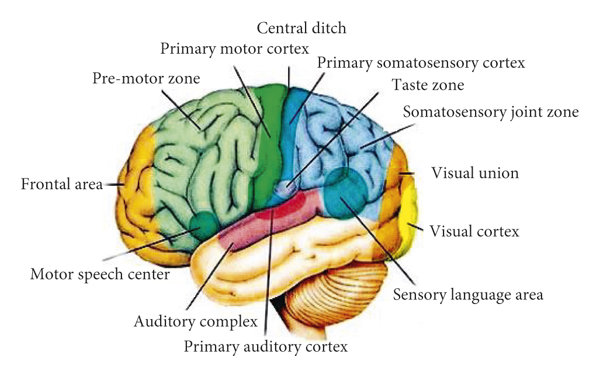}
        \caption{Brain Cortex Map~\citep{zeng2022classifying}}
        \label{fig:1}
    \end{minipage}
\end{figure}
This article presents a comprehensive biofeedback system aiming for two significant high-level objectives, which are stress management and performance enhancement. Meanwhile, in the way of achieving these goals, every part of the system's performance has been optimized and carefully evaluated. In this paper, the brain signals have been recorded using the fNIRS method and utilized in order to classify the stress level of subjects in real-time. For doing so, not only data recording needs to be real-time, but also the pre-processing and processing algorithms operating on brain signals need to be specifically designed to simultaneously have high accuracy and efficiency to be computationally fast in order to be applicable for real-time applications of biofeedback. By using machine-learning algorithms to classify the stress level in real-time, an important online application is required to make a bridge between different parts of the system and create the biofeedback loop by transmitting the ML classifier information to the vibration device. This article would evaluate every part of the system solely and it would further provide a comprehensive evaluation of the overall system to prove the applicability of the system to reduce stress levels and enhance subjects' performances. In order to show that, several propitious tests and qualitative measurements on significant properties of subjects’ performances have been conducted.\newline
Although there are plenty of research studies on brain signal processing, feature extraction \citep{gunawardhane2013non,arefi2018classification,hakimi2018stress}, and stress classification \citep{sundaresan2021evaluating,hakimi2020proposing} in literature, this paper provides novel specialized processing algorithms to make the classification procedure reliable, robust, and accurate while focusing on being applicable in real-time applications. Although deep recurrent neural networks such as RNNs have high accuracy and generalization, they cannot be utilized in this context since they need plenty of training data and they are computationally heavy. Brain signal processing, feature extraction, and classification should be carried out in a period of 2 seconds and with an accuracy of at least 90\%.
Moreover, while it holds true that prior research has delved into stress classification and biofeedback methods \citep{brinkmann2020comparing, hunter2019effect}, a noticeable gap persists in the availability of a comprehensive and robust biofeedback system that addresses all essential components for real-time stress management and performance enhancement. This gap becomes even more apparent when considering the incorporation of a simple yet effective, non-distractive biofeedback procedure. It is evident that existing literature lacks a holistic approach in this regard. Consequently, this paper emerges as a pioneering contribution to designing a biofeedback system that meets the requirements for effective stress management and performance enhancement in real-time.
Paper achievements and contributions can be summarized as follows:
\begin{enumerate}
    \item Comprehensive and automated biofeedback system.
    \item Surprising results of stress management and performance enhancement.
    \item Accurate, robust, and real-time brain signal processing and stress classification.
    \item Simple yet effective fresh idea of the biofeedback vibration device.
    \item Completely real-time software that effectively bridges every part of the system together.
\end{enumerate}

\noindent \noindent The rest of the paper is organized as follows. Section \ref{sec:design} presents the system overview and design in detail. The experimental results are presented in section \ref{sec:ExpRes}. Finally, section \ref{sec:con} concludes the paper.

\section{System Overview and Design}
\label{sec:design}
The overall overview of the biofeedback system can be observed in Fig. \ref{fig:0}. First, the brain signal is recorded and fed into the pre-processing part, which is going to be technically discussed in the signal recording and pre-processing section. Consequently, the filtered signal would be prepared for further evaluation and it then would be processed in order to extract the appropriate features for classification; this part would be explained further in the feature extraction and learning model. Extracted features in the previous step are going to be used firstly for training the learning model first and for each person individually. Then based on the trained model, extracted features in real-time are used for stress classification and measurement in the test phase in which the user is going to fully interact with the tasks. By systematically gauging stress levels during the user's focused engagement with assigned tasks, the system discerns time instances when stress surpasses the typical threshold. During the training phase, a machine learning model dynamically establishes the stress evaluation threshold, reflective of a baseline brain state in non-task-induced scenarios. This preemptive calibration ensures the system's sensitivity to stress deviations specifically during task execution. Upon detecting stress levels exceeding the established threshold, the system promptly notifies the user through a wireless vibration device, heightening awareness of the mental stress state. This deliberate intervention enables subjects to develop conscious and subconscious stress management strategies amid task-related pressures, and foster skill acquisition in stress control while maintaining task accuracy. Consequently, subjects undergo a feedback loop, wherein the interplay between conscious and unconscious stress management contributes to an enhanced ability to navigate stressful situations and uphold task precision. The biofeedback mechanism operates through a dynamic process. While users try to reduce their stress levels, announced to them by biofeedback through a vibration device, their brains gradually go to rest and normal state. Subsequently, the decrement in stress levels aligns with an improvement in overall performance. Notably, as subjects garner positive reinforcement in response to their achievements, a surge in confidence occurs. This heightened confidence, in turn, acts as a catalyst for further stress reduction, creating a continuous and reinforcing cycle over time. Stress level reduction here is caused by reward learning which is a type of reinforcement learning in humans \citep{madan2013toward,dayan2002reward,sander2021reward}. When users get stressed out under the pressure of completing tasks, the vibration sensor starts vibrating as a warning system, then users learn that by reducing the stress level (action) the vibration would stop (reward). Here this learning procedure can be consciously, unconsciously, or in combination of both. The objectives of this paper are to show the effectiveness of the novel idea on performance enhancement and temporary and permanent stress reduction while dealing with daily stressful activities. In order to make the proposed comprehensive system, the following sections are designed and connected efficiently.

\subsection{Applications and Basic Definitions}
The presented system is aimed to be used for people struggling with stress and its side effects as they deal with daily stressful activities. The primary target demographic includes students, office workers, and individuals contending with business-related concerns, as well as those troubled by minor stressors. This system is planned for real-time and portable applications such as for a student learning and solving questions in school or an employee working on routine tasks and assignments in the workplace. As explained earlier, this system will be applied for a specified duration on subjects while they are dealing with these activities to warn the users about their stress levels as biofeedback in the form of vibration. While acknowledging the inherent diversity in stress management and performance enhancement approaches across different subjects, the systematic tracking of main parameters ensures the system's efficacy in fostering stress management abilities over time. Subsequent evaluations, conducted without the aid of biofeedback, serve as a general test for the long-lasting impact of the intervention. If individuals revert to their initial state of mental anxiety, it underscores the necessity of reapplying the system and sustaining the intervention. Overall, it is shown that real-time and portability are the two most important features for the ultimate objectives and they should be taken into consideration for designing the system.
It is worth mentioning that the stress definition in this paper is based on the following. Stress is a relative mental state of the brain \citep{fink2016stress,la2019definition,goodheart2011eating}. If we consider stress level to be zero while people are sleeping, any state of the brain in which subjects are awake and are working on something or thinking about something can be considered stressful since the stress level is higher in those conditions. Therefore, if the rest state of the brain is defined to be stress level zero, as it can be defined relatively, any state of anxiety while dealing with assigned tasks can be described as stress level one. Here the main goal for stress management as will be fully explained in the following parts of the article, is to have a brain-resting mental state while dealing with stressful activities which requires mental business and thinking that enforce stress level one to the subjects. Therefore in this paper, all the classifications and stress management techniques are based on two brain mental state levels which will be referred to as stress levels for the rest of the article.
\begin{figure}[t]
    \centering
    \begin{minipage}{0.49\linewidth}
        \centering
        \includegraphics[width=\linewidth, height=6cm]{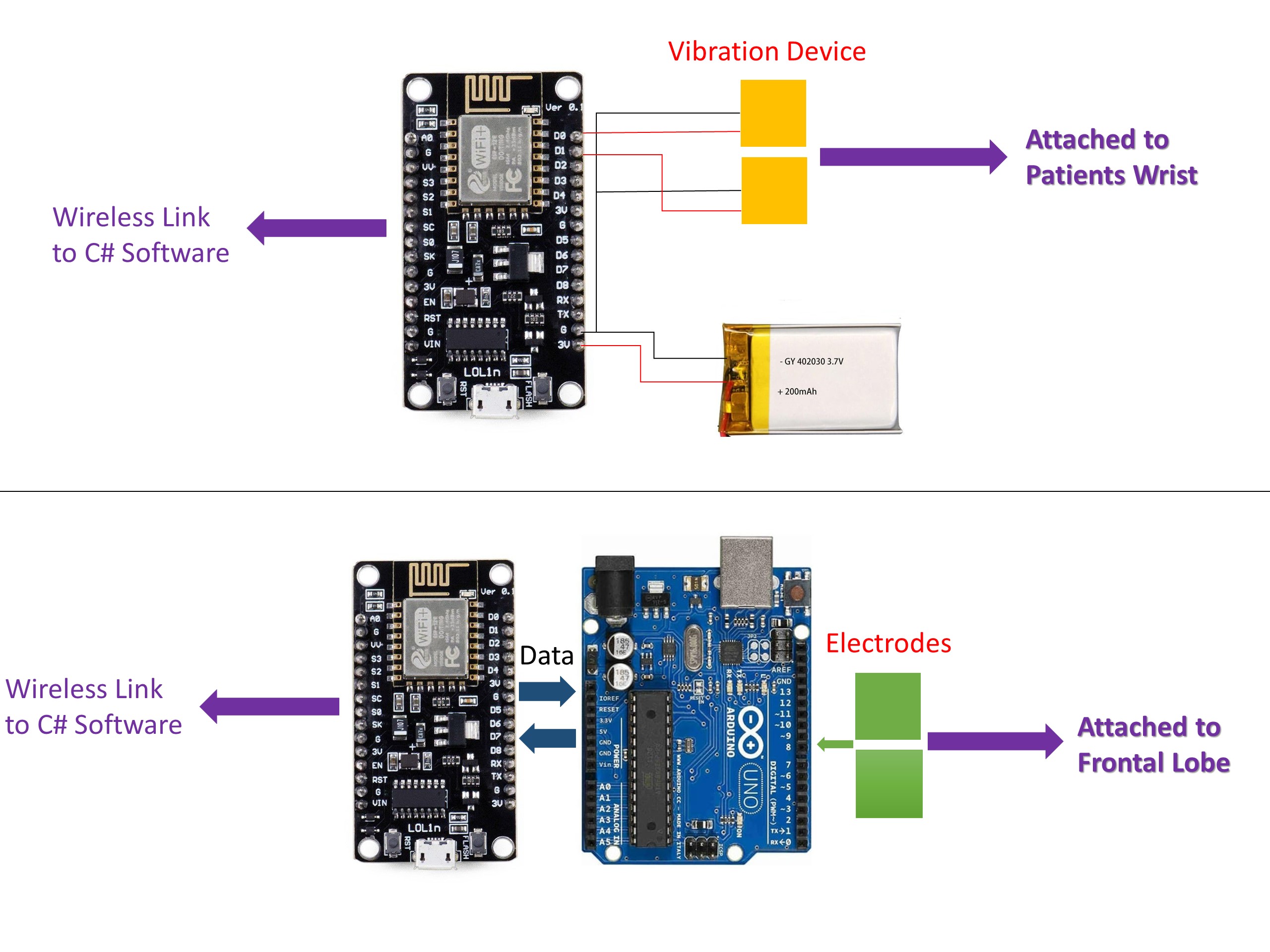}
        \caption{Wireless Brain Signal Recorder and Vibration Device}
        \label{fig:2}
    \end{minipage}%
    \hfill
    \begin{minipage}{0.49\linewidth}
        \centering
        \includegraphics[width=\linewidth, height=4cm]{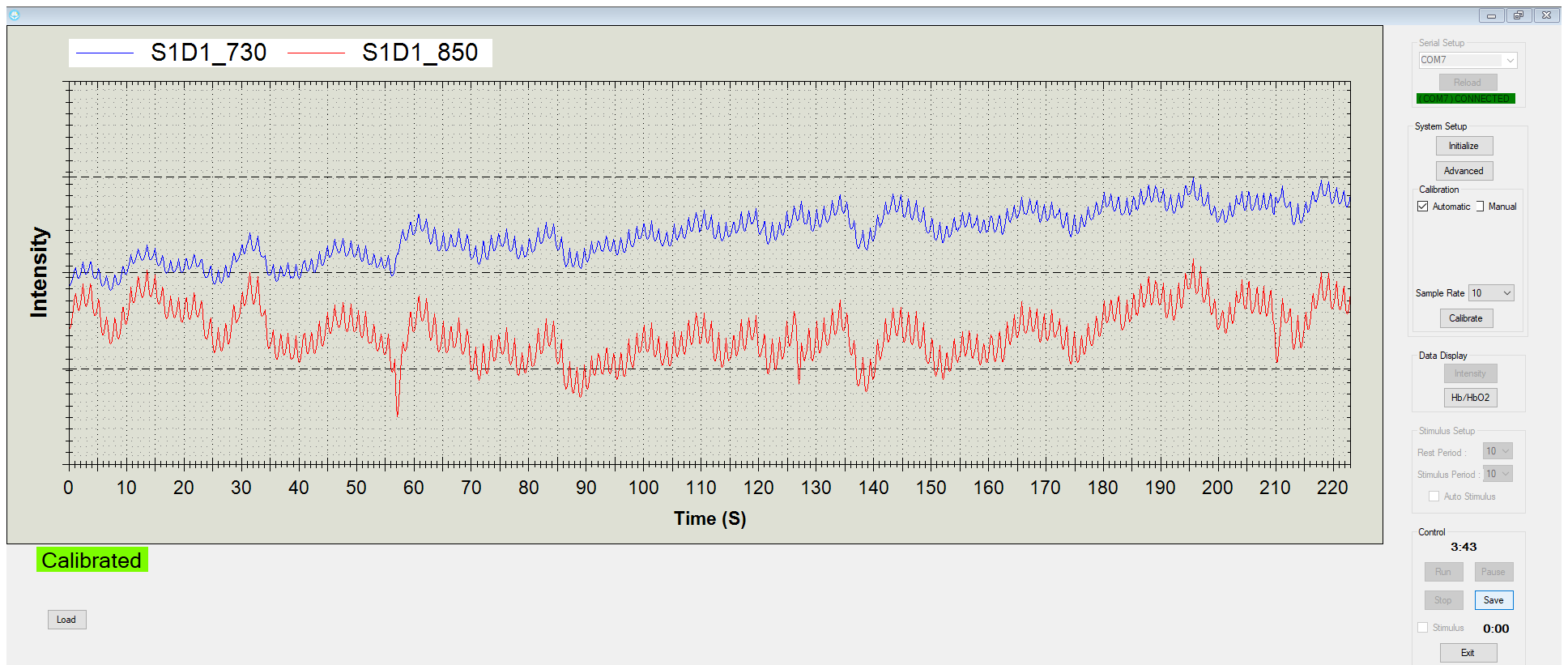}
        \caption{Software GUI and Recorded Brain Signal Data in Real-Time}
        \label{fig:3}
    \end{minipage}
\end{figure}

\subsection{Brain Signal Recording}
In previous studies, it has been proved that different regions of the brain are responsible for different tasks. The brain cortex is divided into four lobes frontal, parietal, temporal, and occipital, each of those responsible for processing different types of information as illustrated in Fig. \ref{fig:1}.  It means when a person is dealing with something in the real world, based on the properties of the task and the type of it, different regions of the brain are activated. For instance, while a person is moving his hands for a specific purpose, the motor cortex is activated as a consequence of that. The motor cortex is in the region of the cerebral cortex involved in the planning, control, and execution of voluntary movement located in the frontal lobe. In addition, sometimes subjects are dealing with complicated tasks, which can activate different parts of the brain simultaneously. It has also been investigated that the prefrontal cortex is responsible for cognitive and emotional tasks and it operates much like a digital computer system, and it is located in the frontal lobe \citep{deco2023one,funahashi2017working,ramnani2004anterior,bloom2009encyclopedia}. Hence, while subjects are solving mathematical questions even those requiring short-term memory, the prefrontal cortex is activated. In this paper, mathematical questions, which will be explained in detail in the evaluation section, are used as the best simple representative for stressful daily activities. This assumption will help us build the foundational framework, which then can be extended to fully complicated tasks. Therefore, it is essential to record the brain signals from the frontal lobe and prefrontal cortex location on the scalp as the representative of the brain digital computer since it contains the information and features dealing with cognitive tasks that can represent the mental stress data. 
There are two different tools for recording brain signals: EEG and fNIRS. fNIRS uses an optical method for recording brain signals. In addition, fNIRS has a better spatial resolution while EEG has a better temporal resolution. Since it is required to discriminate between cortex activities efficiently in this application, fNIRS is the 
 most suitable approach. Moreover, EEG has more electrical noises in the recorded signal and is more susceptible to motion artifacts. Hence, in this paper, Functional Near Infrared Spectroscopy (fNIRS) is incorporated as a method of recording brain signals.
In this paper, two channels for deep and superficial brain signal recording in the prefrontal cortex location of the subjects’ brains are used. In the following plenty of reasons are provided to justify the efficiency and necessity of acquiring two channels for this purpose. Firstly, as mentioned earlier, it is important to make the system portable with the least possible complexity in order to make use of the system in schools and workplaces. In addition, as the second most important feature of the proposed system, stress evaluation and biofeedback loop should operate in real-time. Therefore, it is vital to make the pre-processing and processing phase as simple as possible because enforcing more channels and feature extraction to the system can cause more computational time overheads which causes large delays on portable systems with limited computational resources. Ultimately, the favorable accuracy in stress classification and evaluation can be achieved by two channels of brain signal recording because we are discriminating between two stress levels, and the number of features extracted by these two channels is high enough to get the propitious multi-dimensional feature space for non-linear learning models.

The main hardware for brain signal acquisition is Arduino Uno integrating with an ESP8266 module for wireless data transition to the server for further analysis. ESP and Arduino's codes have been developed in order to make a serial communication between these devices such that initialization parameters and commands to the Arduino are transferred wirelessly to the ESP and it sends the data to the Arduino via serial link. Furthermore, the recorded real-time data can be transmitted to the server via ESP for further processing. Sending and receiving data is based on UDP protocol because it fits the real-time application of sending a stream of data while delay and conflicts are very important. By doing so, the system is enabled to work under mobile real-time applications which does not require a stationary wired setup. Wireless fNIRS brain signal recorder schematic can be seen in Fig. \ref{fig:2}. C\# program for making the GUI for sending the initialization parameters has been developed which can control the calibration parameters and commands along with choosing the number of channels and stop, pause, and run commands. Primarily, as the ESP setup starts working as a virtual access point with a specific name containing fNIRS, the C\# program would look for any Wi-Fi name containing fNIRS and it would list all of the nearby devices from which users can select the desired one. After initializing the link, primary parameters would be transmitted to the device and it would start working and sending back the data just after the calibration procedure. The calibration process calculates baseline parameters such as the environment light. In addition, every IP assignment would be adjusted automatically in order to make the desired wireless networks. Software GUI and visualization of brain signal data are illustrated in Fig. \ref{fig:3}. After running and recording the brain signals, the received data in C\# is written with a specific format into a real-time accessible text file. MATLAB program in real-time would be authorized to read the text file and extract data for further analysis. Hence, the system is enabled to wirelessly send the desired data into a processing framework in real-time. Basically, the main data from the two proposed channels are HBO and HHB concentrations in the brain calculated by the Beer-Lambert formula~\citep{leung2005estimation} (Eq.~\ref{eq:1}) and any additional processing and feature extraction would operate on these four extracted signals in real-time. 
\begin{eqnarray}\label{eq:1}
    A=log(\frac{I}{I_0})=\epsilon \times C \times d \times DPF + G
\end{eqnarray}

\begin{figure}[t]
    \centering
    \begin{minipage}{0.49\linewidth}
        \centering
        \includegraphics[width=\linewidth, height=4cm]{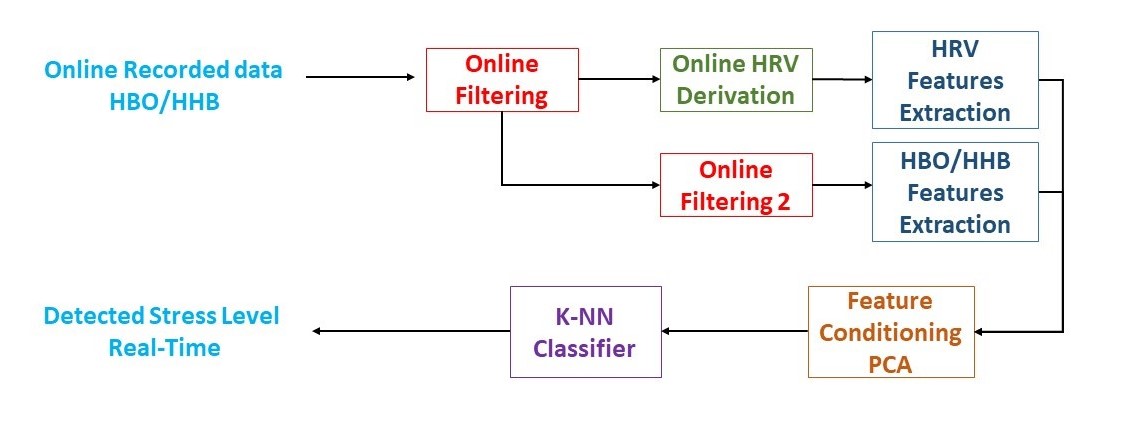}
        \caption{Signal Processing Block Diagram for Feature Extraction in Real-Time}
        \label{fig:4}
    \end{minipage}%
    \hfill
    \begin{minipage}{0.49\linewidth}
        \centering
        \includegraphics[width=\linewidth]{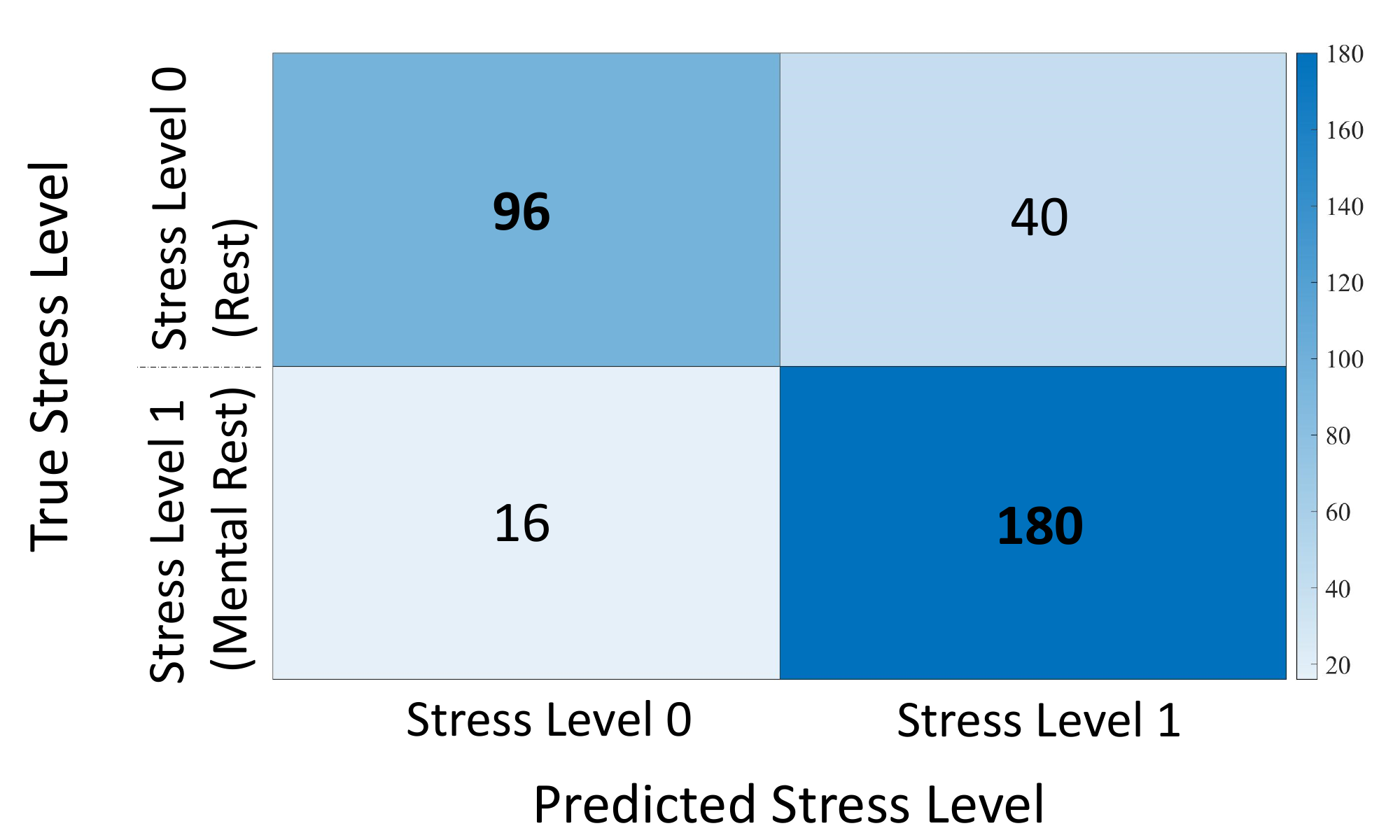}
        \caption{Stress Detection Confusion Matrix}
        \label{fig:stress}
    \end{minipage}
\end{figure}

\subsection{Signal Processing}
The first and most important step is to eliminate noise and artifacts added at the top of the desired signal. It should be taken into consideration that every filter needs to be designed to work in real-time applications. There are indeed different high-quality bandpass filters to eliminate high and low-frequency signals and artifacts but they only work when total derivation of the signal is known. However, in real-time applications, data is being received as time goes by and therefore intuitively the filter should act as a window on top of the current and late indexed received signal data. Therefore, in order to eliminate high-frequency noises of the HBO and HHB signals we should use a real-time low pass filter whose realization is a moving average with tuned window size. A crucial factor to bear in mind concerning the recorded brain signals is the influence of the heartbeat on the HBO and HHB signals, leading to a high-frequency oscillation on top of them. This phenomenon underscores the intricate interplay between cardiovascular activity and the targeted brain signals, introducing a layer of complexity that necessitates careful consideration in the interpretation and analysis of the recorded data. Additionally, this is beneficial since HBO and HHB signal solely can give us HRV information which is essential for stress detection. Therefore, it is required to first extract HRV information and then incorporate real-time low-pass filters to eliminate the effects on top of HBO and HHB signals. The processing procedure is illustrated in Fig. \ref{fig:4}. Ultimately, signals containing the heart rate information are going to be used in HRV real-time measurement using an accurate online peak detection algorithm, and the low pass filtered HBO and HHB signals are going to be filtered again to eliminate the effect of heat rate variability with another fine-tuned moving average.

\begin{figure*}[t]
\centering
\begin{subfigure}[b]{0.45\linewidth}
\includegraphics[width=1\linewidth, height=5cm]{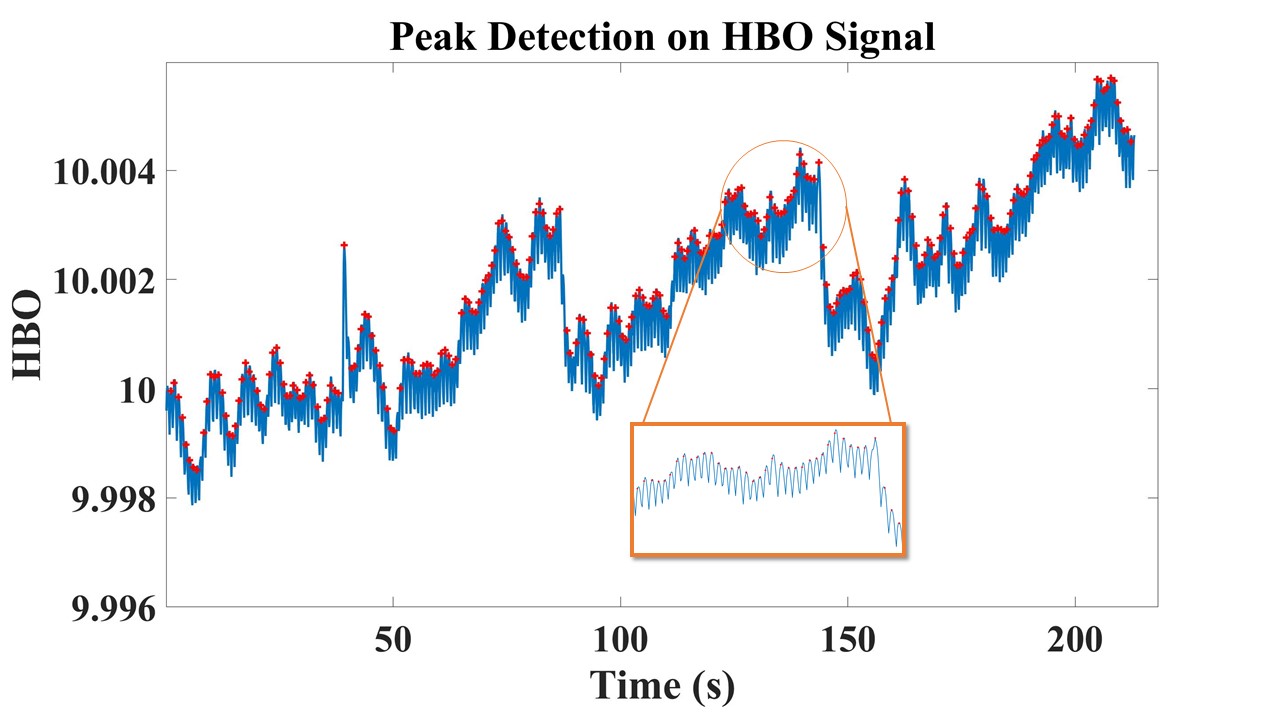} 
\caption{Peak Detection on HBO/HHB Signal}
\label{fig:subimHRV1}
\end{subfigure}
\begin{subfigure}[b]{0.45\linewidth}
\includegraphics[width=1\linewidth, height=5cm]{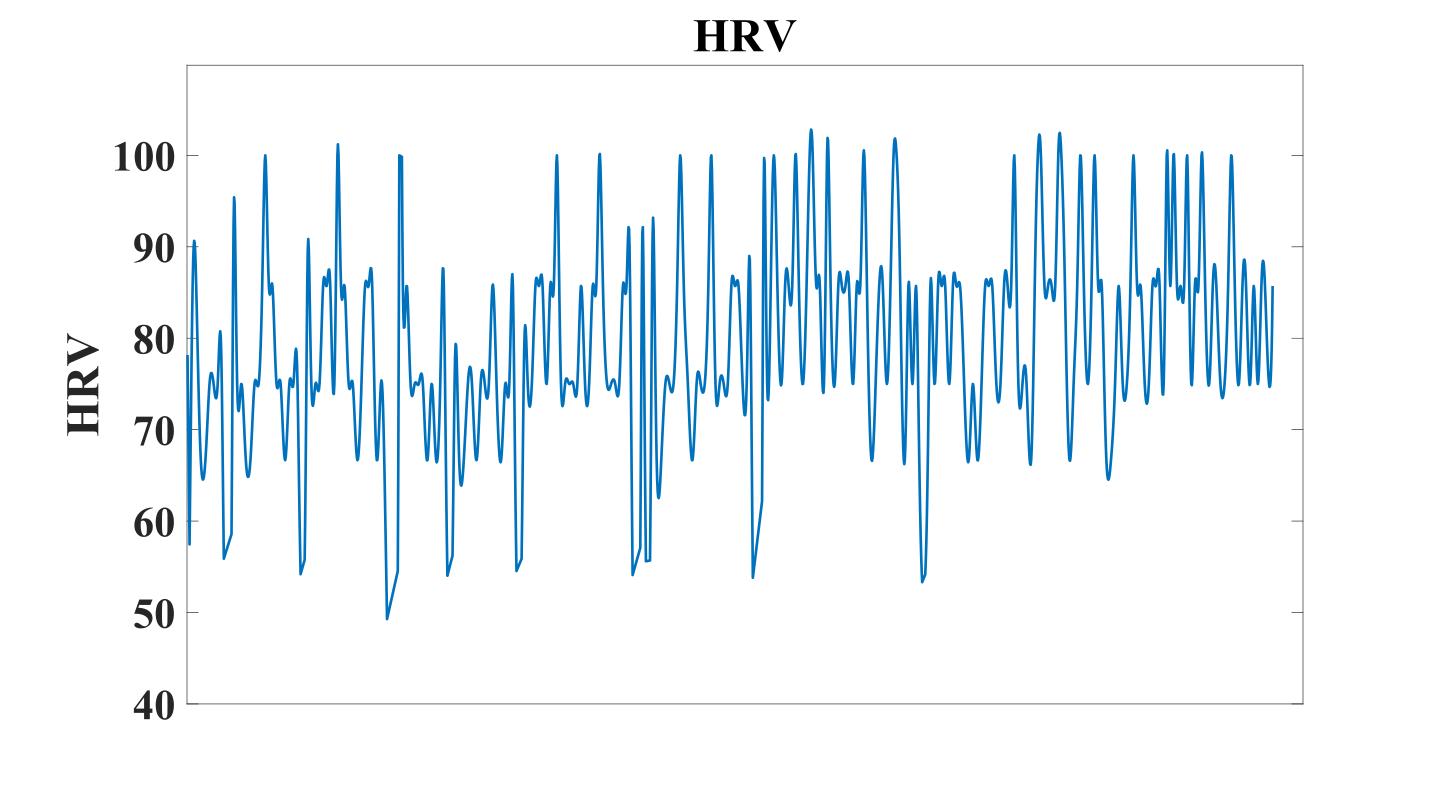}
\caption{Online Heart Rate Variability Derivation}
\label{fig:subimHRV2}
\end{subfigure}
\label{fig:imageHRV}
\caption{}
\end{figure*}

One of the most important features that is going to be used in the learning model is Heart Rate Variability (HRV). In order to extract HRV, which can be simply calculated using the time interval between two adjacent peaks in the heartbeat-containing signal, it is required to first determine the peak time stamps precisely in real-time. There are plenty of real-time peak detection algorithms of which four of them are investigated in \citep{hakimi2018stress}. According to \citep{hakimi2018stress}, the AMPD algorithm is suitable for real-time applications with a mean error of 0.014 and linear correlation of 95.8\%. Hence, in this paper, the AMPD algorithm with two modifications is incorporated to boost the performance and make it robust against the minor motion artifacts. Firstly, one point would be considered as a peak if it is greater than the fine-tuned predetermined threshold while simultaneously the difference between that point and the average calculated value is 10\% greater than the calculated standard deviation as the maximum first difference. Secondly, after detecting a point to be peak based on previous conditions, it should be checked to see whether within a reasonable amount of time 0.25s (based on standard Heat Bit Rate) before this point any peak is detected. If the answer is yes, then this point is a local maximum and it should be eliminated from the list of peak points. Ultimately, the peak points and the corresponding HRV can be derived in real-time based on the AMPD algorithm. An illustration of the peak detection algorithm and HRV derivation proficiency can be observed in Fig. \ref{fig:subimHRV1} and Fig. \ref{fig:subimHRV2} respectively.

\subsection{Feature Extraction}
A summary of all features extracted from the brain signal and calculated HRV, together with their corresponding delays are shown in Table \ref{tab:featues}. The delay periods were adjusted so that the total delay factor of extracted features from HBO and HHB signals be the same. Since the application is aimed to be real-time, we can mostly rely on the time domain features of HBO and HHB signals for classification. However, by calculating HRV based on the time interval between adjacent peaks, the algorithm takes the frequency domain features into consideration. The required delay to calculate the low-order frequency features is not applicable for the real-time purposes of the system. Therefore, the only frequency domain features of the brain signal would correspond to the calculated HRV while other features calculated based on the filtered HBO and HHB signals are in the time domain. Calculated feature space would be extracted based on a propitious window size $N$ operating on incoming data to form new feature space in every time stamp. The time domain features extracted based on the HBO and HHB signal are the mean and standard deviation of the moving average and slope of HHB and HBO which were previously calculated. The frequency domain features based on extracted HRV signal are mean, STD, and maximum HRV along with instantaneous HRV. HRV features can be extracted only based on frequency domain information of the signal which means a more extensive window size $X_3$ is required to capture HRV properties. Consequently, the latency of updating HRV features in real-time would be longer as compared to the delay of the time domain features. Moreover, $X_3$ can be variable based on dynamics and real-time heartbeat variations, and hence it would not be a design parameter. Therefore, the last updated value of HRV features would be evaluated at each time stamp of the classification algorithm. The delay of mean and STD of moving averaged of HBO and HHB consists of $N$ sample delay based on the actual filtering parameter window size and an integer ($X_1$) multiplication of it as the window size for taking the mean or STD of the signal. STD and mean of Slope delay are based on $N$ sample (window size) for filtering the signal, N sample as window size to calculate the slope of the signal, and an integer ($X_2$) multiplication of that just like before. In order to make the delays equal $X_1 = X_2 + 1$ should hold. The choice of $X_2$ should minimize the delay while making the appropriate and reasonable features for the learning model. $X_2 = 1$ would maintain both objectives.

\begin{table*}[t]
\small
  \centering
    \caption{Feature Space Table}
  \label{tab:featues}
\renewcommand{\arraystretch}{2} 
\hspace{-0.4cm}
\begin{tabular}{|>{\centering\arraybackslash}p{1.7cm}|>{\centering\arraybackslash}p{1.7cm}|>{\centering\arraybackslash}p{1.5cm}|>{\centering\arraybackslash}p{1.7cm}|>{\centering\arraybackslash}p{8cm}| }
 \hline
 \multicolumn{5}{|c|}{\textbf{Feature Space}} \\
 \hline
 Feature Name& Original Signal &Feature type & Delay & Derivation\\
 \hline
 Mean Moving Average   & HBO/HHB    &Time Domain&   $N+X_1N$ & $MM_i = \frac{1}{(1+X_1)N}\sum_{n=i-(1+X_1)N} ^{i} MA_n$\\[0.6cm]
 \hline
 STD Moving Average&  HBO/HHB & Time Domain   &$N+X_1N$ & $SM_i = \sqrt{\frac{1}{(1+X_1)N}\sum_{n=i-(1+X_1)N} ^{i} (MA_n - MM_i)^2}$\\[0.6cm]
 \hline
 Mean Slope &HBO/HHB & Time Domain&  $2N+X_2N$ & $MS_i = \frac{1}{(2+X_2)N}\sum_{n=i-(2+X_2)N} ^{i} S_n$\\[0.6cm]
 \hline
 STD Slope    &HBO/HHB & Time Domain& $N+X_1N$ & $SS_i = \sqrt{\frac{1}{(2+X_2)N}\sum_{n=i-(2+X_2)N} ^{i} (S_n - MS_i)^2}$\\[0.6cm]
 \hline
 Mean HRV &   HRV  & Freq. Domain & $X_3N$& $MH_i = \frac{1}{X_3N}\sum_{n=i-X_3N} ^{i} \frac{60\times F_s}{(peak index(n)-(peak index(n-1))}$\\[0.6cm]
 \hline
 STD HRV & HRV  & Freq. Domain &$X_3N$& $SH_i =\sqrt{\frac{1}{X_3N}\sum_{n=i-X_3N} ^{i} (HRV_i - MH_i)^2}$\\[0.6cm]
 \hline
 Max HRV & HRV  & Freq. Domain &$X_3N$& $MXH_i = max(\frac{60\times F_s}{(peak index(n)-(peak index(n-1))})|_{n=i-X_3N}^{i}$\\[0.6cm]
 \hline
 Instantaneous HRV & HRV  & Freq. Domain &$X_3N$ & $IH_i = \frac{60\times F_s}{(peak index(i)-(peak index(i-1))}$\\[0.6cm]
 \hline
\end{tabular}
\end{table*} 

\subsection{Learning Model}
As previously mentioned each of the extracted features is identical in type for both of the channels, and therefore the ultimate feature space includes 20 real-time features. Due to the large number of features and the non-linear complex behavior of the human brain, it is not reasonable to use a linear classifier model for prediction. In this study, we use the K-nearest Neighbour Machine Learning model for stress classification. Moreover, Principal Component Analysis (PCA) is Incorporated for feature conditioning to prune the extracted real-time feature space and eliminate the correlation between features.
In the training phase, a labeled dataset is recorded in a controlled experimental setup. The procedure of gathering labeled data is going to be fully covered in the evaluation section, but as a high-level overview, by assigning repetitive mathematical tasks and rest blocks and setting a specific time interval for each block the labeled data for each participant can be created. This way, it is clear that each recorded raw brain signal data in every time stamp corresponds to either rest or mental business block, which as explained before corresponds to stress levels 0 and 1 respectively. After recording the training dataset by having the subjects do the required assigned tasks in a controlled manner, the KNN model can learn to adjust different hyper-plans in feature space to each stress It must be noted that the training phase for each participant should be done individually and separately since stress level realization and formation are different between people. For instance, for one person stress is realized by changing a specific feature while it can be different in terms of feature type and variation for another person. Meanwhile, the stress classification learning model cannot generalize its properties for each person while training on the limited dataset and different individuals. Therefore, it is crucial to train the model for each person by performing a short primary controlled test which can be seen as a calibration process for the system. In order to test the classification performance of the learning model, the system accuracy has been evaluated to detect the stress level for subjects in a normal controlled condition. we would elaborate on the experimental setup and given tasks in section \ref{sec:ExpRes}, but the overall performance of the stress detection block as an individual part of the system needs to be discussed primarily. In order to have a better understanding of the real-time stress classification model, after recording the brain signal in the training phase and training the model accordingly, the predicted and the ground truth stress levels of the subjects' brains were recorded while they were trying to solve the pre-designed and specified MIST \citep{al2016mental2,dedovic2005montreal} tasks. Furthermore, as the concept of biofeedback and the prominent purpose of the system implies, we have divided the ground truth and predicted stress level into groups of classified stress level samples. Each group would contain 10 samples representing a window size of one second in the time domain. Finally, a binary label will be assigned to each group of recorded data based on the label of majority stress level constituting that. Hence, if the number of stress level 1 is higher in one group the corresponding label for that group would be the mental stress level 1. Fig. \ref{fig:stress} illustrates the confusion matrix of the performed experiment. The accuracy and precision of stress classification is 83\% which is sufficiently high for the purpose of this system while using only two channels of the recorded brain signal using fNIRS. More importantly, $recall$ value for this confusion matrix in this experiment is \textbf{92\%}. This metric is more important than accuracy for the concept of biofeedback since it is more important to detect the True Positive (Mental Stress) accurately. This is to say it is more important to reduce the number of False Negative predicted classes. 

\subsection{Real-Time Stress Classification and Vibration Biofeedback}
After training the model for each individual based on the extracted features and labeled data, we can use it for real-time stress classification. The data recorded from the Arduino are transferred to the C\# program by ESP8266 and it writes it on a real-time read-only accessible text file from which the MATLAB online program extracts new data and processes it with a higher frequency than the recording frequency which would guarantee reading all data in real-time and without any delay. The processing algorithm here means filtering, processing, and feature extraction in real-time. Then, the extracted features are evaluated by the trained K-NN model and the output would be 0 or 1 as the indicator of the stress level.
In the next phase, classified stress levels in real-time in each time stamp would be transmitted to the vibration device by UDP packets containing the stress levels.

\subsection{Vibration Device}
\begin{figure}[t]
    \centering
    \begin{minipage}{0.49\linewidth}
        \centering
        \includegraphics[width=\linewidth, height=6cm]{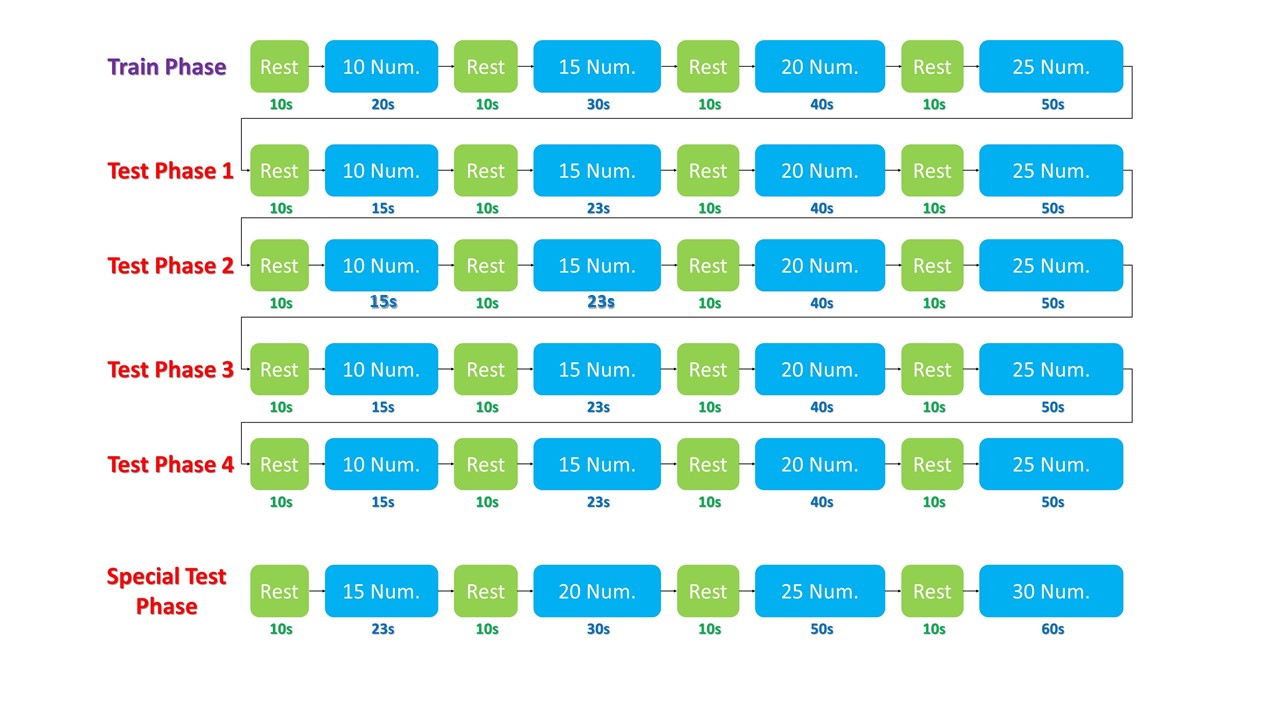}
        \caption{The Block Diagram of Tasks Specified for Mental Stress and Rest Creation}
        \label{fig:tasks}
    \end{minipage}%
    \hfill
    \begin{minipage}{0.49\linewidth}
        \centering
        \includegraphics[width=\linewidth, height=5cm]{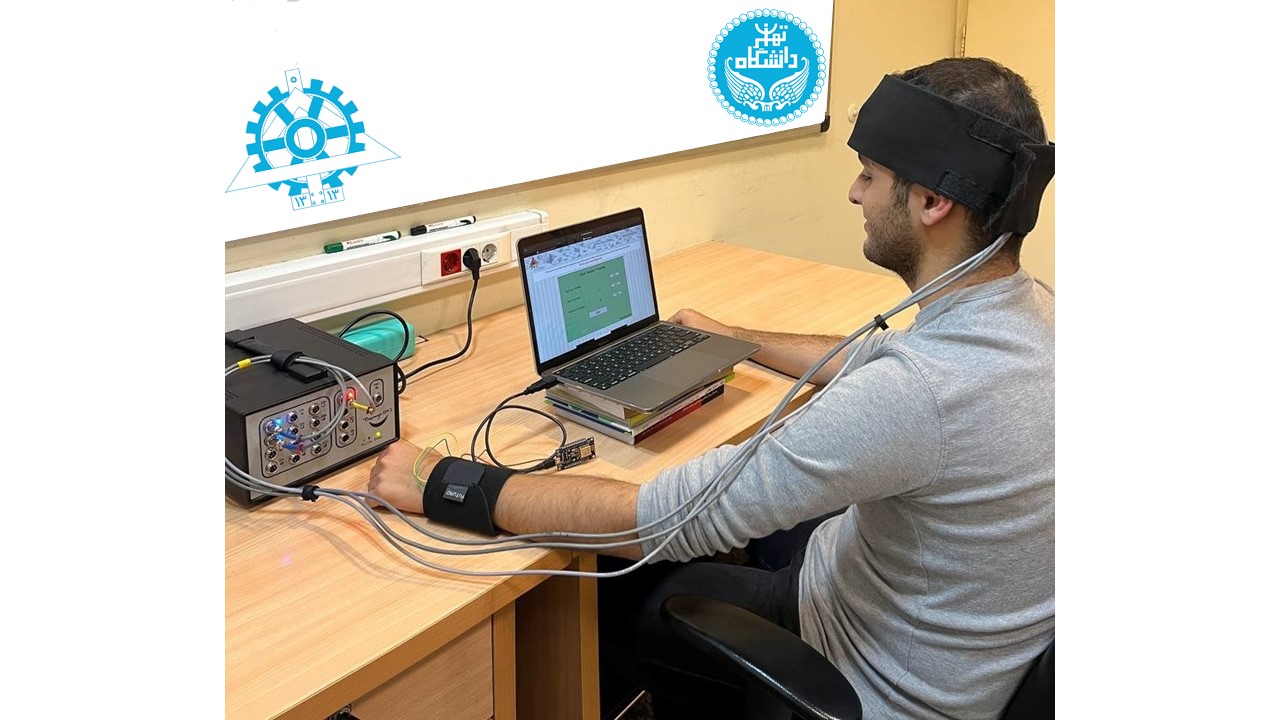}
        \caption{Experimental Setup}
        \label{fig:exp}
    \end{minipage}
\end{figure}

The vibration device contains ESP8266 as the main microcontroller which receives the UDP packets as stress level information wirelessly, one chargeable lithium battery, and one analog voltage controllable vibration module. The code for this microcontroller is developed such that it firstly, connects to the ESP8266 module as the virtual access point on the recording device to create the desired local network containing a brain signal recorder, computer (C\# program), and the vibration device. This is feasible by looking for any network nearby containing fNIRS and predefined codes in the SSID. Secondly, it listens on a predefined port continuously and receives desirable UDP packets. Ultimately, based on the last received packet it generates a voltage on the vibration module port to stimulate the wrist-attached device and create the biofeedback loop. The vibration device should be attached to the subject’s wrist. The vibration system and attachment in the experimental setup can be observed in Fig. \ref{fig:2} and Fig.~\ref{fig:exp} respectively. It is worth mentioning that the vibration level is adjusted not to be unpleasant but just to be in the form of feedback from the stress level. Vibration device design and deployment would naturally filter any spike in the stress classification output making it robust to undesirable sudden changes. All of the design considerations, such as using chargeable batteries for wireless data receiving, etc., have been made to create a portable and compact device for the aforementioned propitious applications.

\section{Biofeedback System Experimental Results}
\label{sec:ExpRes}
In order to show the efficiency of the proposed biofeedback system, different experimental setups have been designed. It is essential to divide the subjects into two main groups in which individuals deal with assigned tasks with and without the vibration biofeedback device. Hence, the first group of participants would do the tasks while the biofeedback vibration device has been attached to their wrist and they are aware of their mental stress level in real-time. The second group of subjects would participate in the experiments in a normal condition without any biofeedback. This way we would be able to evaluate the performance of each group in accuracy enhancement and stress reduction in the period of doing the tasks and compare the results to conclude the efficiency of the proposed system. Moreover, as it has been discussed in the system overview and design section, it is required to evaluate the efficiency of the stress detection part, which is going to be tested in another form of experiment to have control over the mental stress level of subjects for the third group of participants. The first group contains 15 participants (8 men and 7 women) from different backgrounds while all of them are students. Similarly, the second group has been chosen from students with various fields of study, containing 3 women and 2 men. Also, the stress detection part has been tested on one separate participant in order to avoid having a bias of doing the experiment carried on to system experiments. The average age of subjects participating in the tests is 25.

\subsection{Experimental Procedure}
In the beginning, all subjects were asked to do two essential mental tests ("ASRS" and "SAIN") to make sure that they are not suffering from mental health problems such as ADHD, OCD, etc. Next, participants were asked to sit in a comfortable chair in a silent, ventilated, and free from any movement and environmental distraction factors in a laboratory environment. Meanwhile, subjects were asked to list five desirable objects, places, or names that make them calm whenever they think about them. Then, before starting the experiments individuals were allowed to sit 5 minutes quietly in order to compensate for stressing parameters existing in an experimental procedure. The experiments and the calibration phase were carried out in a dark room. This needs to be done in order to eliminate all the illumination noises in the environment while using Functional Near-infrared Spectroscopy for recording the brain signals. The experimental setup is illustrated in Fig. \ref{fig:exp}. The experiment has an ethics code of "IR.UT.SPORT.REC.1402.114" from the Iranian University of Medical Sciences.

\begin{figure*}[t]
\centering
\begin{subfigure}{0.57\linewidth}
\includegraphics[width=1\linewidth, height=6cm]{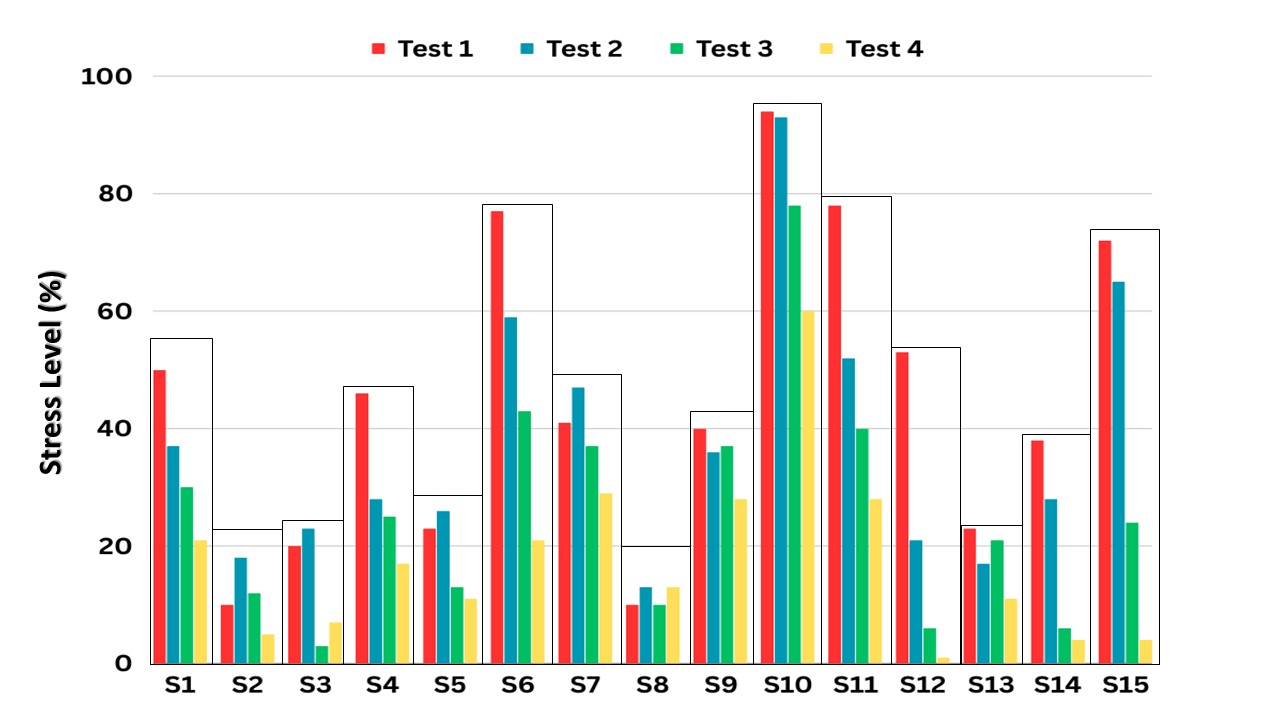} 
\caption{With Vibration Biofeedback}
\label{fig:subimstressw}
\end{subfigure}
\begin{subfigure}{0.42\linewidth}
\includegraphics[width=1\linewidth, height=6cm]{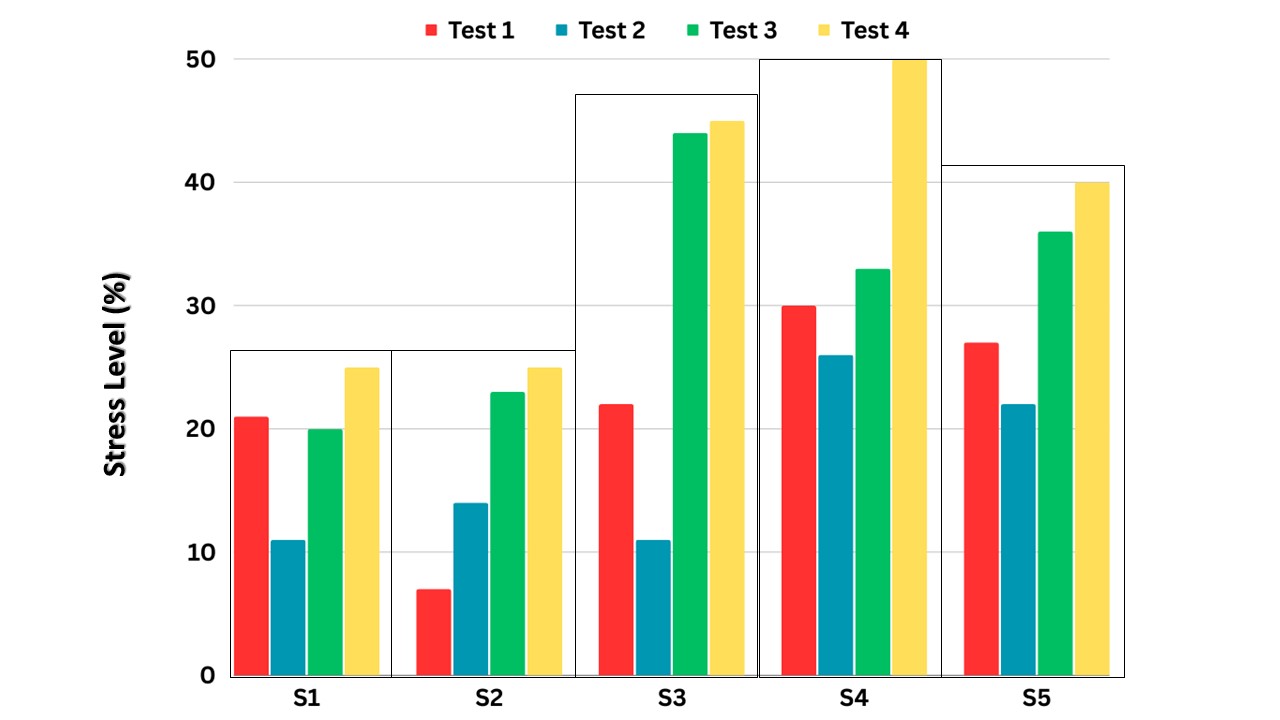}
\caption{Without Vibration Biofeedback}
\label{fig:subimstresswo}
\end{subfigure}
\label{fig:imagestress}
\caption{Stress Level Reduction Evaluation in Various Phases of Experiment for Different Subjects}
\end{figure*}

\subsection{Task sequence and Experimental Conditions}
Tasks in this experiment, inspired by MIST \citep{al2016mental2,dedovic2005montreal}, are designed to induce mental stress in the subjects. In order to make the mental stress level consistent throughout the experiment, participants perform simple mathematical calculation which is only summation but in different modalities. Task sequences and instructions for the first two groups are the same, and an overview of that is illustrated in Fig. \ref{fig:tasks}. Primarily instructions on the task sequences were given to the subjects to be familiar with the experimental procedure. Then, in order to train the learning model for each participant, we designed the training phase in the task sequence. This phase consists of two major parts rest and calculation which are responsible for inducing rest and mental stress respectively. In the rest block subjects are asked to remain relaxed and just think about one of the listed items for 10 seconds. On the other hand, in the calculation blocks participants are asked to calculate the summation of numbers displayed on the screen. In each calculation block, numbers remained on the monitor for 2 seconds, and the total of displayed numbers in each separate calculation block was 10, 15, 20, and 25 in turn. After finishing displaying numbers in each calculation block the brain signal recording was paused in order to eliminate the effect of displaying results stress on subjects. Finally, at the end of the training phase, the learning model would be trained based on the recorded data, and ready to detect the stress level for specific individuals. It is worth mentioning that after each completion phase (training or test phases) subjects were asked to take a break and relax for 5 minutes. \newline
After training the model, patients were asked to do four consecutive test phases with a similar construction to the training phase. The only difference in the test phase is that numbers in the first two blocks of calculation remain on the screen for 1.5 seconds instead of 2 seconds. The first group of subjects with a vibration device attached to their wrist were kept aware of their mental stress level by utilizing vibration biofeedback and the detected stress level by learning mode. Hence, participants were asked to try to minimize their stress levels by thinking about one of the listed items which makes them calm whenever the vibration device alarms a high stress level. Simultaneously, subjects were asked to focus on solving the mathematical questions while dealing with stress reduction. On the other hand, the second group of participants were asked to do the exact same test phases, but without any biofeedback. It is worth mentioning that in order to minimize the effect of strong backgrounds on mathematics a special test phase has been designed, details of that can be observed in Fig. \ref{fig:tasks}. If subjects answer all the mathematical questions in calculation blocks perfectly, they are assigned to the special test in the next coming test phase. Finally, in order to test the stress detection part of the system, participants were assigned to do the training phase repeatedly 5 times.

\begin{figure*}[t]
\centering
\begin{subfigure}{0.45\linewidth}
\includegraphics[width=1\linewidth, height=4cm]{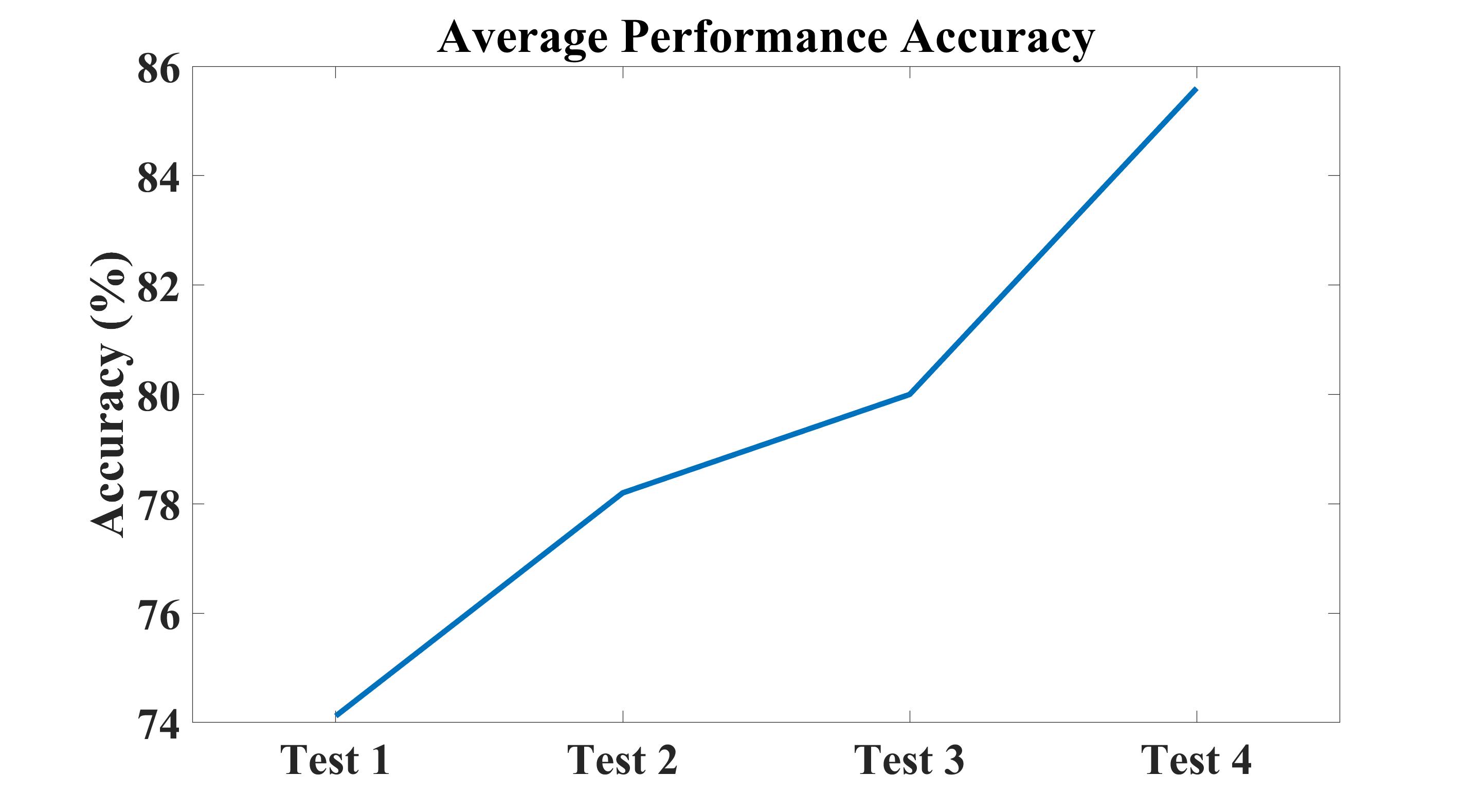} 
\caption{With Vibration Biofeedback}
\label{fig:subimperw}
\end{subfigure}
\begin{subfigure}{0.45\linewidth}
\includegraphics[width=1\linewidth, height=4cm]{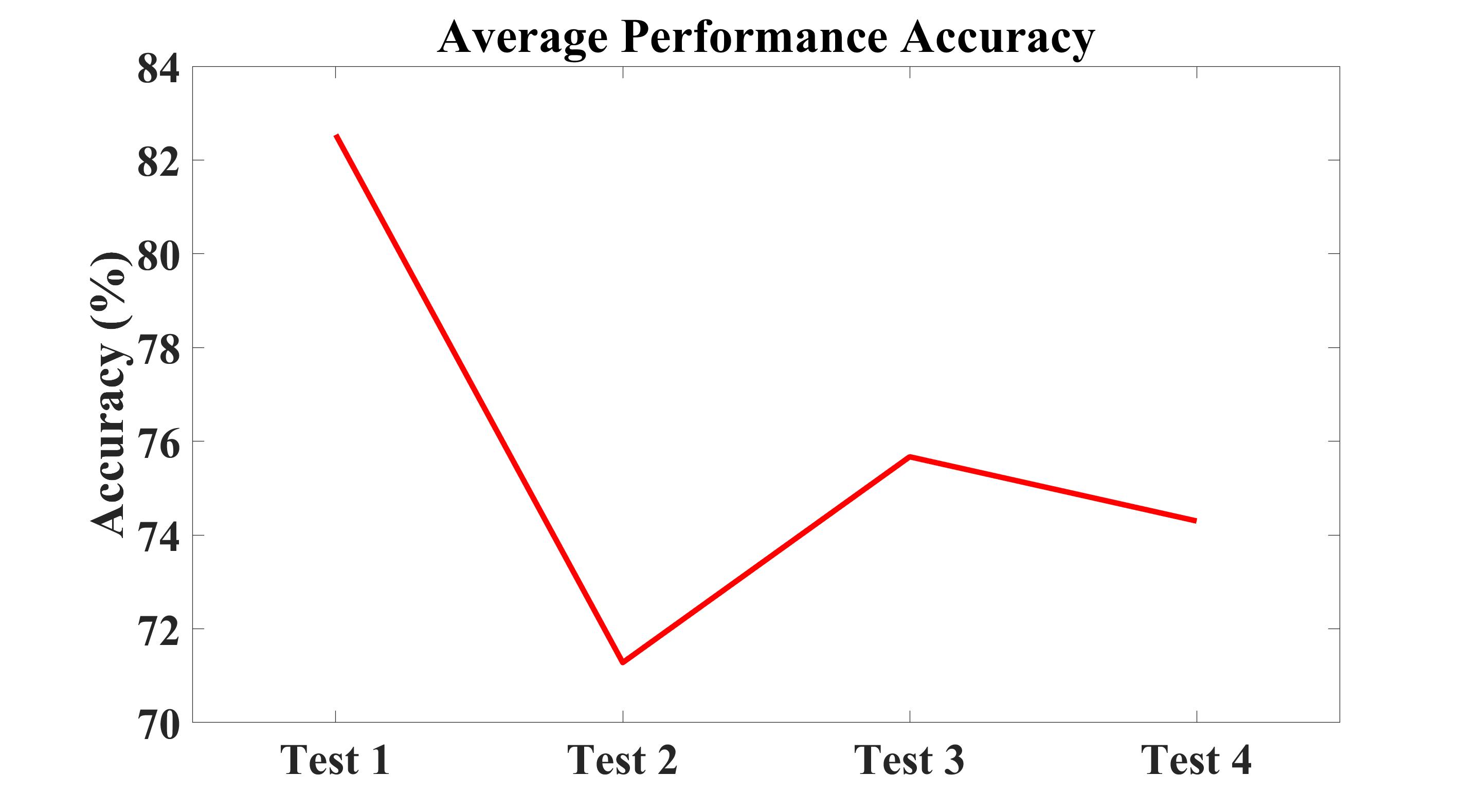}
\caption{Without Vibration Biofeedback}
\label{fig:subimperwo}
\end{subfigure}
\label{fig:imageper}
\caption{Average Performance (Accuracy) Enhancement Evaluation in Various Phases of Experiment}
\end{figure*}

\begin{table*}[htbp]
\small
\centering
\caption{Result Summary}
\label{tab:results}
\begin{tabular}{ |>{\centering\arraybackslash}p{2.5cm}|>{\centering\arraybackslash}p{3cm}|>{\centering\arraybackslash}p{3.5cm}|>{\centering\arraybackslash}p{3cm}|>{\centering\arraybackslash}p{3.5cm}| }
\hline
\multicolumn{5}{|c|}{\textbf{Stress Reduction and Performance Enhancement}} \\
\hline
Phase &  Stress Level Reduction (With Biofeedback) & Performance Enhancement (With Biofeedback) & Stress Level Reduction (Without Biofeedback) & Performance Enhancement (Without Biofeedback)\\
\hline
Test2 - Test1  &16\% & 7.1\%   & 6&-13.7 \\
\hline
Test3 - Test2 &33\% & 3.9\% & -60\% & 3.9\%\\
\hline
Test4 - Test3    &25\%& 8.3\%&  -42\%& -1.8\%\\
\hline
Total Evaluation & 55\% & 15.5\% & -90\% & -9\%\\
\hline
\end{tabular}
\end{table*}

\subsection{Results and Discussion}
    The data in each time stamp of the experiment has been recorded carefully for all participants in order to show the efficiency of the proposed system. This data consists of subjects' stress level through time, the accuracy of answering mathematical questions, and all other primary extracted features such as HRV, HBO/HHB signals, etc. As can be seen from Fig. \ref{fig:subimstressw} and Fig. \ref{fig:subimstresswo} representing the average percentage of stress level of subjects in the different test phases for participants with biofeedback and without that respectively. It is clear from the bar graph that subjects with the biofeedback vibration device experienced a downward trend in the average mental stress level as they proceeded to the next test phases. For the majority of subjects that used our system during the experiment, substantial stress management and reduction throughout the total period of the test can be clearly observed. On the other hand, not only the subjects' stress levels do not experience a downward trend in the normal condition, but also it is clear that the stress level increased in the final phase as compared to the initial phase. Moreover, for some subjects, it is evident that stress levels rose consistently which shows the efficiency of our system for stress management as compared to the ordinary situation.\newline 
Regarding the performance enhancement in the subjects’ answers, the average accuracy of responses in each test phase has been calculated for all participants. The average accuracy of subjects' performances using the biofeedback system and those without it, are illustrated in Fig. \ref{fig:subimperw} and Fig. \ref{fig:subimperwo} in turn. Clearly, these figures prove the applicability of our proposed system to enhance subjects’ performances since the response accuracy of participants with vibration devices increased over time significantly in comparison with that of individuals without the biofeedback system. Surprisingly, the performance of subjects without vibration devices decreased while at first it was expected to remain constant.\newline
A summary of participant performances in stress reduction and accuracy enhancement is reported in Table \ref{tab:results}. The average normalized value of stress reduction in percentage between consequent test phases and in total for both groups has been summarized in this Table \ref{tab:results}. It suggests that stress reduction for the group of participants with biofeedback in three transition phases and in total is 16\%, 33\%, 25\%, and 55\% respectively. On the other hand, the reported average value for the other group without our proposed systems is 6\%, -60\%, -42\%, and -90\% in turn. Moreover, the average accuracy enhancement of subjects with vibration device is 7.1\%, 3.9\%, and 8.3\% in subsequent test phases, and 15.5\% in total, while that of participants without biofeedback is -13.7\%, 3.9\%, and -1.8\% in subsequent test phases, and -9\% in total. Overall, this table strongly illustrates the efficiency of our proposed system over ordinary conditions. \newline
An important takeaway from the results can be deducted which is biofeedback in the form of vibration can efficiently help subjects with stress management and performance enhancement. Primarily, the performance enhancement is significant even at the very first phases of the experiment which shows the prominent role of biofeedback. On the other hand, stress reduction in the first phases is relatively low as compared to that in the consequent test phases. This phenomenon suggests that at the first stage, subjects are trying to find the trade-off to increase their performance in both stress management and accuracy. Although it is hard for them to stay focused on the tasks while they should try to manage their stress level, they firstly focus on stress reduction in the cost of losing some accuracy enhancement. As it is evident from the result stress reduction increases substantially at the second stage to 33\% while accuracy enhancement drops to 3.9\%. However, in the next stage subjects learn to control their stress level and make it consistent with a very minor drop to 25\% while maintaining the accuracy increasing it back to 8.3\%. At the final stage, participants unconsciously learn to maintain stress reduction and performance enhancement simultaneously. However, it is clear that with more trials, subjects are expected to have better results in performance enhancement, but even in a limited period of four tests, the performance enhancement result is substantially promising. It is worth mentioning that test phase duration can be optimized for different individuals, but for the purpose of this paper, it was required to make it consistent for all subjects. Overall, results suggest 55\% stress reduction and 24.5\% (15.5\% – (-9\%)) performance enhancement while using our proposed system.

\section{Conclusions and Future Work}
\label{sec:con}
This article presented a comprehensive biofeedback system for stress reduction and performance enhancement for individuals dealing with stressful daily activities. It has been mentioned that the fNIRS methodology has been incorporated into the brain signal recording device, and a specified wireless vibration deceive has been utilized to create the biofeedback loop. The system has been designed with a special focus to make the system portable and work in real-time. This paper described signal processing and feature extraction parts of the system, and it explained and demonstrated the structure and efficiency of the stress classification part. Furthermore, we have tested our system as an integrity under standardized experimental conditions, and we have highlighted the advantages of our system for stress management and performance enhancement. The proposed system has been tested on two separate groups of participants from different backgrounds with specialized task sequences in order to show the efficiency of the vibration biofeedback system. The findings revealed that our system was able to help subjects reduce their stress level by 55\% and increase their accuracy by 24.5\%. Overall, the proposed system was successfully helpful for participants regarding stress reduction and performance enhancement, which from the knowledge of authors is the first of its kind to present that concept.\newline
For future research directions, we propose exploring the potential applications of biofeedback systems in more intricate scenarios that require the recording and analysis of brain signals from diverse regions of the cerebral cortex. In this study, we introduced a comprehensive system for real-time assessment of mental stress levels, aiding users in managing stressful tasks and improving their performance. This adaptable system is primed for updates to accommodate more complex activities and mitigate induced anxiety by integrating the established biofeedback framework. Such advancements would necessitate advanced brain signal recording equipment and learning models capable of capturing the effects of various complicated activities on brain signals in real-time.\newline

\section*{Acknowledgment}
Our special thanks go to Doctor Rostami, who provided constructive suggestions on the Stress Management side of the system design.
The authors would also like to thank all the people who participated
in the study, including subjects and students who collaborated.

\bibliographystyle{unsrtnat}
\bibliography{main}  






\end{document}